\documentclass[pre,amsmath,amssymb,superscriptaddress]{revtex4}

\usepackage{amsmath,amssymb}
\usepackage[usenames]{color}
\usepackage[dvipsnames]{xcolor}
\usepackage{amssymb}
\usepackage{dsfont}
\usepackage{grffile}
\usepackage{amsmath}
\usepackage{amstext}
\usepackage{amssymb}
\usepackage{textcomp}
\usepackage{xspace}
\DeclareSymbolFontAlphabet{\amsmathbb}{AMSb}

\newcommand{\brasil}{Departamento de F\'{\i}sica--Instituto de Ci\^{e}ncias Exatas, Universidade Federal de Minas Gerais, CP 702, 30.161-970 Belo Horizonte MG, Brazil}

\begin{document}

\title{Transport in boundary-driven quantum spin systems: One-way street for the energy current}

\author{Deborah Oliveira}
\author{Emmanuel Pereira}
\email{emmanuel@fisica.ufmg.br}
\affiliation{\brasil}
\author{Humberto C. F. Lemos}
\email{humbertolemos@ufsj.edu.br}
\affiliation{Departamento de Estat\'{\i}stica, F\'{\i}sica e Matem\'atica, CAP --
Universidade Federal de S\~ao Jo\~ao del-Rei, 36.420-000, Ouro Branco, MG, Brazil}

\begin{abstract}
We study transport properties in boundary-driven asymmetric quantum spin chains
given by $\mathit{XXZ}$ and $\mathit{XXX}$ Heisenberg models. Our approach exploits symmetry
transformations in the Lindblad master equation associated to the dynamics of
the systems. We describe the mathematical steps to build the unitary transformations related to the symmetry properties. For general target
polarizations, we show the occurrence of the one-way street phenomenon for
the energy current, namely, the energy current does not change in magnitude
and direction as we invert the baths at the boundaries. We also analyze
the spin current in some situations, and we prove the uniqueness of the
steady state for all investigated cases. Our results, involving nontrivial
properties of the energy flow, shall interest researchers working on
the control and manipulation of quantum transport.
\end{abstract}
%\pacs{67.57.Lm, 03.65.Yz, 75.10.Jm, 67.80.−s}

%67.57.Lm Spin dynamics
%67.57.Hi Transport properties
%67.80.−s Quantum solids
%75.10.Jm Quantized spin models, including quantum spin frustration
% 75.40.Gb Dynamic properties (dynamic susceptibility, spin waves, spin diffusion, dynamic scaling, etc.)
% 75.76.+j Spin transport effects
% 03.65.Yz	Decoherence; open systems; quantum statistical methods

\maketitle

\section{Introduction}

A bedrock of nonequilibrium statistical physics is the understanding of the transport laws \cite{LLP, Dhar, BLiRMP}. In particular, the study of the energy flow properties is of theoretical and experimental interest: a good
example is the investigation of thermal rectification. Motivated by the amazing progress of modern electronics due to the invention of transistor, electric diode and other nonlinear solid state devices, several works are
devoted to the investigations of asymmetries in the energy current in order to build thermal diodes \cite{Casati+, Prapid10}, a device in which the magnitude of the energy current changes as we invert the system between two baths.

A subject of increasing attention nowadays is the study of such transport laws in the quantum regime. Stimulated by the emerging field of quantum thermodynamics, by the development of nanotechnology and the possibility of
experimental manipulation of small quantum systems, the study of quantum models becomes mandatory.

Quantum spin chains, in specific, are exhaustively investigated. They are the archetypal models of open quantum systems, and are related to problems in several different areas: condensed matter, cold atoms, optics, quantum
information, etc. Their boundary-driven versions, i.e., systems with target polarization at the boundaries are recurrently studied \cite{Prosen+, Z, Kar, GL1, GL2}. The energy current of these boundary-driven  systems,
differently of the weakly coupled models, usually involves heat and work \cite{P18, G-NJP, FB}. It is an important information: when we ignore the work component, incorrect conclusions may be obtained \cite{G-NJP, Levy-EPL}.

The present article addresses the investigation of some (a)symmetries in the energy current of boundary-driven Heisenberg ($\mathit{XXX}$) and $\mathit{XXZ}$ models. In specific, we show the occurrence of the one-way street phenomenon for the energy current to several asymmetric Heisenberg and $\mathit{XXZ}$ with general cases of different boundary polarizations.
 Such a phenomenon means that the energy current is the same as we invert the baths at the boundaries, that is, it does not
change in magnitude and direction. Thus, the phenomenon is, in some way, related to (but stronger than) rectification.

It is important to emphasize that, as said, the energy current is not only heat, and so, there is no thermodynamic inconsistency in the occurrence of the one-way street effect. For more details, see Refs.\cite{P18, G-NJP, FB}.

The dynamics associated to the models, as usual, is given by a Lindblad master equation (LME) \cite{BP}. To establish our results we exploit symmetries of the density matrix and of the LME. And these results are independent of the
system size and of the transport regime.

The existence of the one-way street phenomenon was established in Ref.\cite{SPL} by a direct computation of the steady density matrix and the energy current for a $\mathit{XXZ}$ model with $\sigma^{z}$ polarization at the boundaries.
The argument of symmetries appeared in Ref.\cite{Prapid} for the same case, and in a recent letter \cite{EPL} we stated, without presenting a mathematical proof, the possibility of a ubiquitous occurrence of such phenomenon
for systems with general target spin polarization at the boundaries. In the present paper, we give the mathematical proofs for the energy current property; we also show that, in some cases, the spin current changes the sign
as we invert the baths, differently of the energy flow. And, an important
mathematical point, we prove the uniqueness of the steady distributions for
the cases treated here.

The rest of the paper is organized as follows. In section 2, we introduce the
model and describe the approach. In section 3, we analyze several cases of
different target polarizations and present the mathematical steps. In section
4, we prove the uniqueness of the steady states. Section 5 is devoted to the
final remarks.

\section{Models and approach}

Now we introduce the models to be treated here, the LME, the approach to be used and some previous results.

We consider here standard quantum spin models, namely, the $\mathit{XXZ}$ and Heisenberg ($\mathit{XXX}$) chains. For the Hamiltonian of the asymmetric version of the spin 1/2 $\mathit{XXZ}$ chain, we take
\begin{equation}
\mathcal{H} =  \sum_{i=1}^{N-1}\left\{ \alpha\left(\sigma_{i}^{x}\sigma_{i+1}^{x} + \sigma_{i}^{y}\sigma_{i+1}^{y}\right)  + \Delta_{i}\sigma_{i}^{z}\sigma_{i+1}^{z} \right\}
 ~, \label{hamiltonian}
\end{equation}
where $\sigma_{i}^{\beta}$ ($\beta = x, y, z$) are the Pauli matrices. We are interested in cases involving asymmetric distributions for the anisotropy parameter $\Delta_{i}$, for example, a graded distribution:
$\Delta_{1}<\Delta_{2}<\cdots<\Delta_{N-1}$.

For the Heisenberg model, we take the Hamiltonian
\begin{equation}
\mathcal{H} =  \sum_{i=1}^{N-1} \alpha_{i}\left(\sigma_{i}^{x}\sigma_{i+1}^{x} + \sigma_{i}^{y}\sigma_{i+1}^{y}  + \sigma_{i}^{z}\sigma_{i+1}^{z} \right) ~, \label{hamiltonian2}
\end{equation}
where $\alpha_{i}$ is asymmetrically distributed.

The open quantum systems to be analyzed are given by the steady states of the LME
\begin{equation}
\frac{d\rho}{d t} = i[\rho, \mathcal{H}] + \mathcal{L}(\rho) ~,\label{master}
\end{equation}
where we assume $\hbar =1$,  $\rho$ is the density matrix, the dissipator $\mathcal{L}(\rho)$  describes the coupling with the baths and it is given by
\begin{eqnarray}
\mathcal{L}(\rho) &=& \mathcal{L}_{L}(\rho) + \mathcal{L}_{R}(\rho) ~, \nonumber\\
\mathcal{L}_{L,R}(\rho) &=& \sum_{s=\pm} L_{s}\rho L_{s}^{\dagger} -
\frac{1}{2}\left\{ L_{s}^{\dagger}L_{s} , \rho \right\} ~,\label{dissipator}
\end{eqnarray}
$\{\cdot,\cdot\}$ above describes the anti-commutator; different $L_{s}$ will be specified later.

The spin and the energy current are derived from the LME and continuity equations, see, e.g., Ref.\cite{Mendoza-A} for details. For the $\mathit{XXZ}$ chain,
the spin current is
\begin{equation}
\langle J^{M}_{j} \rangle = 2\alpha \langle \sigma_{j}^{x} \sigma_{j+1}^{y} - \sigma_{j}^{y}\sigma_{j+1}^{x} \rangle ~.
\end{equation}
Adding in the Hamiltonian the interaction with an external magnetic field
$\sum_{j=1}^{N} B_{j}\sigma_{j}^{z}$, the energy current becomes
\begin{eqnarray}
\langle J^{E}_{j}\rangle &=& \langle J_{j}^{\mathit{XXZ}}\rangle + \langle J_{j}^{M}\rangle ~,\nonumber\\
\langle J_{j}^{\mathit{XXZ}} \rangle &=&  2\alpha \langle \alpha \left( \sigma_{j-1}^{y}\sigma_{j}^{z} \sigma_{j+1}^{x} - \sigma_{j-1}^{x}\sigma_{j}^{z}\sigma_{j+1}^{y}\right) \nonumber\\
&& + \Delta_{j-1,j}\left( \sigma_{j-1}^{z}\sigma_{j}^{x} \sigma_{j+1}^{y} - \sigma_{j-1}^{z}\sigma_{j}^{y}\sigma_{j+1}^{x}\right) \nonumber\\
&& + \Delta_{j,j+1}\left( \sigma_{j-1}^{x}\sigma_{j}^{y} \sigma_{j+1}^{z} - \sigma_{j-1}^{y}\sigma_{j}^{x}\sigma_{j+1}^{z}\right)\rangle  ~,\nonumber\\
\langle J_{j}^{M} \rangle &=& \frac{1}{2} B_{j}\langle J_{j-1} + J_{j}\rangle ~. \label{currents}
\end{eqnarray}

It is important to recall that there is a remarkable difference between
symmetric and asymmetric $\mathit{XXZ}$ chains. For the symmetric case we have
$\langle J_{j}^{\mathit{XXZ}} \rangle = 0$ \cite{Mendoza-A}. And so, the energy
current becomes proportional to the spin current, and vanishes as $B=0$. But
it does not follow in the asymmetric case as shown, by a direct computation,
in Ref.\cite{SPL} for a system with $\sigma^{z}$ polarization at the boundaries.

Turning to the Heisenberg Hamiltonian, the expressions for the currents
become
\begin{equation}
\langle J^{M}_{j} \rangle = 2\alpha_{j} \langle \sigma_{j}^{x} \sigma_{j+1}^{y} - \sigma_{j}^{y}\sigma_{j+1}^{x} \rangle ~,
\end{equation}
\begin{eqnarray}
\langle J_{j}^{\mathit{XXZ}} \rangle &=&  2\alpha_{i-1} \alpha_{i}\langle \left( \sigma_{j-1}^{y}\sigma_{j}^{z} \sigma_{j+1}^{x} - \sigma_{j-1}^{x}\sigma_{j}^{z}\sigma_{j+1}^{y}\right) \nonumber\\
&& + \left( \sigma_{j-1}^{z}\sigma_{j}^{x} \sigma_{j+1}^{y} - \sigma_{j-1}^{z}\sigma_{j}^{y}\sigma_{j+1}^{x}\right) \nonumber\\
&& + \left( \sigma_{j-1}^{x}\sigma_{j}^{y} \sigma_{j+1}^{z} - \sigma_{j-1}^{y}\sigma_{j}^{x}\sigma_{j+1}^{z}\right)\rangle  ~.\nonumber\\
%\langle F_{j}^{B} \rangle &=& \frac{1}{2} B_{j}\langle J_{j-1} + J_{j}\rangle ~.
\label{currents2}
\end{eqnarray}

We now describe our strategy to prove the current properties, in particular,
the one-way street phenomenon. In some way, we follow Popkov and Livi \cite{PopLi}.
We exploit symmetries in the LME to show that, if
$\rho$ is a steady state solution of the LME, then there is a unitary transformation $U$ (to be built) such that $U\rho U^{\dagger}$ is a solution
of the LME with inverted baths. By uniqueness (to be proved), it is the
steady state with inverted baths. Then we turn to the energy current and
show that the average with the new steady state is the same of that with
the initial steady state. That is, the energy current does not change as we
invert the baths: this is the one-way street phenomenon.

To be precise, in the steady state the LME becomes
$$
0 = -i[\mathcal{H}, \rho] + \mathcal{L}(\rho) ~.
$$
It means that, in order to perform our analysis, we must find a unitary transformation $U$ such that
\begin{equation}
\mathcal{H} = U\mathcal{H}U^{\dagger}, ~~~~ \mathcal{L}_{\rm inv. baths}\left(U\rho U^{\dagger}\right) = U \mathcal{L} U^{\dagger} ~.
\end{equation}
Moreover, to show the one-way street phenomenon we need to prove that
$U J U^{\dagger} = J$.

In the next section, we find $U$ for several different dissipators, i.e.,
several different boundary polarizations, such that these relations are
satisfied.

%%%%%%%%%%%%%%%%%%%%%%%%%%%%%%%%%%%%%%%%%%%%%%%%%%%%%%%%%%%%%%%%%%%%%%%%%%%%%%%%%%%%%%%%%%%%%%%%%%%%%%%%%%%%%%%%%%%%%%%%%%%%%%%%%%%%%%%%%%%%%%%%%%%%%%%%%%%%%%%%%%%%%%%%%%%%%%%%%%%%%%%%%%%%%%%%%%%%%%%%%%%%%%%%%%%

%%%%%%%%%%%%%%%%%%%%%%%%%%%%%%%%%%%%%%%%%%%%%%%%%%%%%%%%%%%%%%%%%%%%%%%%%%%%%%%%%%%%%%%%%%%%%%%%%%%%%%%%%%%%%%%%%%%%%%%%%%%%%%%%%%%%%%%%%%%%%%%%%%%%%%%%%%%%

\section{Unitary transformations and symmetry results}

Now we will build the unitary transformations in order to exploit the symmetries of the Lindblad equations and prove some current properties, in particular, the one-way street phenomenon for the energy current.

We begin by noting that any unitary matrix can be written as (the reader
can prove it)

\begin{equation}U=
\begin{pmatrix}
a & b \\
- e^{i\varphi}b^* & e^{i\varphi}a^*
\end{pmatrix}~,
\end{equation}
where $a$, $b \in \mathbb{C}$, $\varphi \in \mathbb{R}$
 and $|a|^2 + |b|^2 =1$.

Then, we analyze several cases involving different boundary polarizations. We also investigate two different graded systems: the two first cases are related to the $\mathit{XXZ}$ chain, and the other ones to Heisenberg model.
First we take the case in which the polarization is in the $x$ direction in one boundary, and in some generic angle in the plane $xy$ in the other
boundary. Precisely, we consider the Lindblad operators

%\section{X-XY}

\begin{align}
& K_{+}^{L} = \sqrt{\gamma(1+f)} \left(\frac{\sigma_{1}^{y} + i\sigma_{1}^{z}}{2}\right)~, \nonumber\\
& K_{-}^{L} = \sqrt{\gamma(1-f)} \left(\frac{\sigma_{1}^{y} - i\sigma_{1}^{z}}{2}\right)~, \nonumber\\
& K_{+}^{R} = \sqrt{\gamma(1-f)}\left(\frac{\cos \theta \sigma_{N}^{x} + \sin \theta \sigma_{N}^{y}+ i\sigma_{N}^{z}}{2}\right)~, \nonumber\\
& K_{-}^{R} = \sqrt{\gamma(1+f)}\left(\frac{\cos \theta \sigma_{N}^{x} + \sin \theta \sigma_{N}^{y}- i\sigma_{N}^{z}}{2}\right)~,
\end{align}
where $\gamma$ is the coupling constant and $f$ is the driving strength (we
take $f_{L}=-f_{R}=f$).

To perform the change between the baths, we need to find a unitary operator
such that
\begin{align}
&U(\sigma^{y} + i\sigma^{z})U^{\dagger} = \cos \theta \sigma^{x} + \sin \theta \sigma^{y}- i\sigma^{z}~, \nonumber\\
&U(\sigma^{y} - i\sigma^{z})U^{\dagger} = \cos \theta \sigma^{x} + \sin \theta \sigma^{y}+ i\sigma^{z}~, \nonumber\\
&U (\cos\theta \sigma^{x} + \sin \theta \sigma^{y}+ i\sigma^{z})U^{\dagger} = \sigma^{y} - i\sigma^{z}~, \nonumber\\
&U (\cos \theta \sigma^{x} + \sin \theta \sigma^{y}- i\sigma^{z})U^{\dagger} = \sigma^{y} + i\sigma^{z}~.
\end{align}
We may still have factors such as $-1$, $i$ or $-i$ on the right hand side
without any further problem.

Given such conditions, we see that it is enough to find a unitary matrix
 $A$ such that the operation $A$ $({\cdot})$ $A^{\dagger}$ transforms as:
\begin{gather*}
\sigma^{y} \xrightarrow[\text{(1)}]{\text{}} \cos\theta\sigma^{x} + \sin{\theta}\sigma^{y} \xrightarrow[\text{(2)}]{\text{}} \sigma^{y}\\
\sigma^{z} \xrightarrow[\text{(3)}]{\text{}} -\sigma^{z}
\end{gather*}
And so,  $U$ will be given by
\begin{equation}
U = A \otimes A \otimes ... \otimes A ~.
\end{equation}

Carrying out the computation:
\begin{equation}
\begin{split}
A \sigma^{z} A^{\dagger} &= \begin{pmatrix}
a & b \\
-e^{i\varphi}b^* & e^{i\varphi}a^* \end{pmatrix} \begin{pmatrix}
1 & 0 \\
0 & -1
\end{pmatrix}A^{\dagger} \\
%&= \begin{pmatrix}
%a & -b \\
%-e^{i\varphi}b^* & -e^{i\varphi}a^* \end{pmatrix}\begin{pmatrix}
%a^* & -e^{-i\varphi}b \\
%b^* & e^{-i\varphi}a
%\end{pmatrix} \\
& = \begin{pmatrix}\label{eq4.20}
|a|^2-|b|^2 & -e^{-i\varphi}ab - e^{-i\varphi}ab \\
-e^{i\varphi}b^*a^* - e^{i\varphi}b^*a^* & |b|^2 - |a|^2
\end{pmatrix}~.
\end{split}
\end{equation}
But we want
\begin{equation}\label{eq4.21}
A \sigma^{z} A^{\dagger} = -\sigma^{z} = \begin{pmatrix}
-1 & 0\\
0 & 1
\end{pmatrix}~.
\end{equation}
It implies that $|b|^2 - |a|^2 = 1$. As we already have $|b|^2 + |a|^2 = 1$, then $|a|^2 = 0$, and so, $a=0$, $|b|^2 = 1$.

Now the matrix $A$ is given by
\begin{equation}
A = \begin{pmatrix}
0 & b \\
-e^{i\varphi}b^* & 0
\end{pmatrix}~.
\end{equation}

To find $b$ we perform the computation
\begin{equation}
\begin{split}
A\sigma^{y}A^{\dagger} &= \begin{pmatrix}
0 & b\\
-e^{i\varphi}b^* & 0
\end{pmatrix}\begin{pmatrix}
0 & -i\\
i & 0
\end{pmatrix}A^{\dagger} \\
%&= \begin{pmatrix}
%ib & 0\\
%0 & ie^{i\varphi}b^*
%\end{pmatrix} \begin{pmatrix}
%0 & -e^{-i\varphi}b\\
%b^* & 0
%\end{pmatrix}\\
&= \begin{pmatrix}
0 & -ie^{-i\varphi}b^2\\
ie^{i\varphi}b^{*2} & 0
\end{pmatrix}~.
\end{split}
\end{equation}

We want
\begin{equation}\label{eq4.24}
A\sigma^{y}A^{\dagger} = \cos{\theta}\sigma^{x} + \sin{\theta}\sigma^{y} = \begin{pmatrix}
0 & e^{-i\theta}\\
e^{i\theta} & 0
\end{pmatrix}~,
\end{equation}
we take $\varphi = \theta$, and it leads us to $-ib^2=1$. Then, it is enough
to take  $b=\frac{1+i}{\sqrt{2}}$.
%\begin{equation*}
%b^2 = \left(\frac{1+i}{\sqrt{2}} \right)\left(\frac{1+i}{\sqrt{2}} \right) = \frac{1+i+i-1}{2} = i = \frac{1}{-i}~.
%\end{equation*}
Hence,
\begin{equation}
A = \frac{1}{\sqrt{2}}\begin{pmatrix}
	0 & 1+i\\
	-e^{i\theta}(1-i) & 0
	\end{pmatrix}
\end{equation}
is the desired unitary matrix.

Carrying out some computation we find that
\begin{equation*}
A(\cos{\theta}\sigma^{x}+\sin{\theta}\sigma^{y})A^{\dagger} = \sigma^{y}~.
\end{equation*}

Now we analyze the $\mathit{XXZ}$ Hamiltonian
\begin{equation*}
H = \sum_{i=1}^{N-1} \alpha(\sigma_{i}^{x}\sigma_{i+1}^{x}+\sigma_{i}^{y}\sigma_{i+1}^{y}) + \Delta_{i}\sigma_{i}^{z}\sigma_{i+1}^{z}~.
\end{equation*}
First, we note that
\begin{equation}\label{eq4.22}
A\sigma^{x}A^{\dagger} = -\sin{\theta}\sigma^{x} + \cos{\theta}\sigma^{y}~.
\end{equation}
Then, it follows that
\begin{equation}
\begin{split}
UHU^{\dagger} &= \sum_{i=1}^{N-1} \alpha(A\sigma_{i}^{x}A^{\dagger}A\sigma_{i+1}^{x}A^{\dagger} + A\sigma_{i}^{y}A^{\dagger}A\sigma_{i+1}^{y}A^{\dagger}) \\
&+ \Delta_{i}A\sigma_{i}^{z}A^{\dagger}A\sigma_{i+1}^{z}A^{\dagger}\\
&= \ldots\\
&= \sum_{i=1}^{N-1} \alpha(\sigma_{i}^{x}\sigma_{i+1}^{x}+\sigma_{i}^{y}\sigma_{i+1}^{y}) + \Delta_{i}\sigma_{i}^{z}\sigma_{i+1}^{z}~.
\end{split}
\end{equation}
That is, $UHU^{\dagger} = H$.

Before investigating the effect of $U$ on the currents, we note that
\begin{equation*}
A^{\dagger}\sigma^{x}A = -\sin{\theta}\sigma^{x} + \cos{\theta}\sigma^{y}~,
\end{equation*}
\begin{equation*}
A^{\dagger}\sigma^{y}A =\cos{\theta}\sigma^{x}+ \sin{\theta}\sigma^y~,
\end{equation*}
\begin{equation*}
A^{\dagger}\sigma^{z}A =    -\sigma^{z}~.
\end{equation*}

The energy current is
\begin{equation}
\begin{split}
\hat{J}^{E} = 2\alpha[\alpha(\sigma_{i-1}^y\sigma_{i}^z\sigma_{i+1}^{x} - \sigma_{i-1}^{x}\sigma_{i}^{z}\sigma_{i+1}^{y}) \\ + \Delta_{i-1}(\sigma_{i-1}^z\sigma_{i}^{x}\sigma_{i+1}^y - \sigma_{i-1}^{z}\sigma_{i}^{y}\sigma_{i+1}^{x}) \\ +\Delta_{i}(\sigma_{i-1}^{x}\sigma_{i}^{y}\sigma_{i+1}^{z} - \sigma_{i-1}^{y}\sigma_{i}^{x}\sigma_{i+1}^{z}) ]~,
\end{split}
\end{equation}
and so the effect of $U$ is
\begin{equation}
\begin{split}
U^{\dagger}\hat{J}^{E}U = 2\alpha\{\alpha[(\cos{\theta}\sigma_{i-1}^{x} + \sin{\theta}\sigma_{i-1}^{y})(-\sigma_{i}^{z})(-\sin{\theta}\sigma_{i+1}^{x} + \cos{\theta}\sigma_{i+1}^{y})\\-(-\sin{\theta}\sigma_{i-1}^{x}+\cos{\theta}\sigma_{i-1}^{y})(-\sigma_{i}^{z})(\cos{\theta}\sigma_{i+1}^{x} + \sin{\theta}\sigma_{i+1}^{y})] \\
+ \Delta_{i-1}[(-\sigma_{i-1}^{z})(-\sin{\theta}\sigma_{i}^{x} + \cos{\theta}\sigma_{i}^{y})(\cos{\theta}\sigma_{i+1}^{x} + \sin{\theta}\sigma_{i+1}^{y})\\-(-\sigma_{i-1}^{z})(\cos{\theta}\sigma_{i}^{x} + \sin{\theta}\sigma_{i}^{y})(-\sin{\theta}\sigma_{i+1}^{x} + \cos{\theta}\sigma_{i+1}^{y})] \\
+ \Delta_{i}[(-\sin{\theta}\sigma_{i-1}^{x}+\cos{\theta}\sigma_{i-1}^{y})(\cos{\theta}\sigma_{i}^{x}+\sin{\theta}\sigma_{i}^{y})(-\sigma_{i+1}^{z})\\-(\cos{\theta}\sigma_{i-1}^{x}+\sin{\theta}\sigma_{i-1}^{y})(-\sin{\theta}\sigma_{i}^{x} + \cos{\theta}\sigma_{i}^{y})(-\sigma_{i+1}^{z})]\}\\
=\ldots = \hat{J}^{E}~.
\end{split}
\end{equation}
It shows the occurrence of the one-way street phenomenon: the energy current
is the same, it keeps the same value and direction as we invert the reservoirs at the boundaries.

Taking the spin current
\begin{equation}
\hat{J}^M = 2\alpha(\sigma_{i}^{x}\sigma_{i+1}^{y} -\sigma_{i}^{y}\sigma_{i+1}^{x}) ~,
\end{equation}
the effect of $U$ is
\begin{equation}
\begin{split}
U^{\dagger}\hat{J}^{M}U &= 2\alpha[A^{\dagger}\sigma_{i}^{x}AA^{\dagger}\sigma_{i+1}^{y}A - A^{\dagger}\sigma_{i}^{y}AA^{\dagger}\sigma_{i+1}^{x}A]\\
&=2\alpha[(-\sin{\theta}\sigma_{i}^{x} + \cos{\theta}\sigma_{i}^{y})(\cos{\theta}\sigma_{i+1}^{x} + \sin{\theta}\sigma_{i+1}^{y})\\&-(\cos{\theta}\sigma_{i}^{x}+\sin{\theta}\sigma_{i}^{y})(-\sin{\theta}\sigma_{i+1}^{x}+\cos{\theta}\sigma_{i+1}^{y})]\\
&= 2\alpha[-\sin{\theta}\cos{\theta}\sigma_{i}^{x}\sigma_{i+1}^{x} -\sin^2{\theta}\sigma_{i}^{x}\sigma_{i+1}^{y} + \cos^2{\theta}\sigma_{i}^{y}\sigma_{i+1}^{x} \\ &+ \sin{\theta}\cos{\theta}\sigma_{i}^{y}\sigma_{i+1}^{y} + \sin{\theta}\cos{\theta}\sigma_{i}^{x}\sigma_{i+1}^{x} - \cos^2{\theta}\sigma_{i}^{x}\sigma_{i+1}^{y} \\& + \sin^2{\theta}\sigma_{i}^{y}\sigma_{i+1}^{x} - \sin{\theta}\cos{\theta}\sigma_{i}^{y}\sigma_{i+1}^{y}] \\
&= 2\alpha(\sigma_{i}^{y}\sigma_{i+1}^{x}-\sigma_{i}^{x}\sigma_{i+1}^{y})~.
\end{split}
\end{equation}

That is,
\begin{equation}
U^{\dagger}\hat{J}^{M}U = -\hat{J}^M~,
\end{equation}
in other words, the spin current keeps the value and inverts the direction as
we invert the reservoirs at the boundaries; there is no spin rectification, no further effect.

{\bf X-Y orthogonal polarization.} Let us consider the set of Lindblad operators for one boundary
\begin{equation}
\begin{split}
L_{1} &= \alpha(\sigma_{1}^{x}+i\sigma_{1}^{y})~, ~~~~L_{2} = \beta(\sigma_{1}^{x}-i\sigma_{1}^{y})~,\\
V_{1} &= p(\sigma_{1}^{y} +i\sigma_{1}^{z})~, ~~~~V_{2} = q(\sigma_{1}^{y}-i\sigma_{i}^{z})~,\\
W_{1} &= u(\sigma_{1}^{z} + i\sigma_{1}^{x})~, ~~~~W_{2} = v(\sigma_{1}^{z} - i\sigma_{1}^{x})~.
\end{split}
\end{equation}
And, for the other boundary,
\begin{equation}
\begin{split}
L_{3} &= \beta(\sigma_{N}^{x} + i\sigma_{N}^{y})~, ~~~~ L_{4} = \alpha(\sigma_{N}^{x} - i\sigma_{N}^{y})~,\\
V_{3} &= v(\sigma_{N}^{y} + i\sigma_{N}^{z})~, ~~~~ V_{4} = u(\sigma_{N}^{y} - i\sigma_{N}^{z})~,\\
W_{3} &= q(\sigma_{N}^{z} + i\sigma_{N}^{x})~, ~~~~ W_{4} = p(\sigma_{N}^{z} - i\sigma_{N}^{x})~.
\end{split}
\end{equation}

For this case, the inversion of the baths can be given by a unitary operator
 $U = A \otimes A \otimes ... \otimes A$ such that
\begin{equation}
A\sigma^{x}A^{\dagger} = -\sigma^{y}~, ~~~~ A\sigma^{y}A^{\dagger} = -\sigma^{x}~,~~~~ A\sigma^{z}A^{\dagger} = -\sigma^{z}~.
\end{equation}
Indeed, with such an operator we have the transformations
\begin{equation}
\begin{split}
L_{1} &\rightarrow -iL_{4}~,~~~~L_{2} \rightarrow iL_{3}~,~~~~
V_{1} \rightarrow -iW_{4}~,~~~~ V_{2} \rightarrow iW_{3}~,\\
W_{1} &\rightarrow -iV_{4}~,~~~~ W_{2} \rightarrow iV_{3}~,~~~~
L_{3} \rightarrow -iL_{2}~,~~~~ L_{4} \rightarrow iL_{1}~,\\
V_{3} &\rightarrow -iW_{2}~,~~~~ V_{4} \rightarrow iW_{1}~,~~~~
W_{3} \rightarrow -iV_{2}~,~~~~ W_{4} \rightarrow iV_{1}~.
\end{split}
\end{equation}

We will use the general representation for a matrix $A\in SU(2)$:
\begin{equation} A = \begin{pmatrix}
a_{r}+ia_{i} & b_{r}+ib_{i}\\
-b_{r}+ib_{i} & a_{r}-ia_{i}
\end{pmatrix}~,
\end{equation}
where $a_{r},a_{i},b_{r}$ and $b_{i} \in \mathbb{R}$ e $a_{r}^2 + a_{i}^2 + b_{r}^2 + b_{i}^2 = 1$.

Turning to the computations,
\begin{equation}
\begin{split}
A\sigma^{x}A^{\dagger} &= \begin{pmatrix}
a_{r}+ia_{i} & b_{r}+ib_{i}\\
-b_{r}+ib_{i} & a_{r}-ia_{i}
\end{pmatrix}\begin{pmatrix}
0 & 1\\
1 & 0
\end{pmatrix}A^{\dagger}\\
%&=\begin{pmatrix}
%b_{r}+ib_{i} & a_{r}+ia_{i}\\
%a_{r}-ia_{i} & -b_{r}+ib_{i}
%\end{pmatrix}\begin{pmatrix}
%a_{r}-ia_{i} & -b_{r}-ib_{i}\\
%b_{r}-ib_{i} & a_{r}+ia_{i}
%\end{pmatrix}\\
&=\begin{pmatrix}
c_{11} & c_{12}\\
c_{21} & c_{22}
\end{pmatrix}~,
\end{split}
\end{equation}
where
\begin{equation}
\begin{split}
c_{11} &= 2(a_{r}b_{r}+a_{i}b_{i})~,\\
c_{12} &= a_{r}^2 + b_{i}^2-a_{i}^2-b_{r}^2 + 2(a_{i}a_{r}-b_{i}b_{r})i~,\\
c_{21} &= a_{r}^2+b_{i}^2-a_{i}^2-b_{r}^2 +2(b_{i}b_{r}-a_{i}a_{r})i~,\\
c_{22} &= -2(a_{r}b_{r} + a_{i}b_{i})~.
\end{split}
\end{equation}
As we want
\begin{equation}A\sigma^{x}A^{\dagger} =\begin{pmatrix}
0 & i\\
-i & 0
\end{pmatrix} ~,
\end{equation}
we must have
\begin{equation}\label{eq4.50}
\begin{split}
a_{r}b_{r} + a_{i}b_{i} &= 0~,\\
a_{r}^2 - a_{i}^2 + b_{i}^2 - b_{r}^2 &= 0~,\\
a_{i}a_{r} - b_{i}b_{r} &= \frac{1}{2}~.
\end{split}
\end{equation}

For the $\sigma^{z}$ transformation
\begin{equation}
\begin{split}
A\sigma^{z}A^{\dagger} &= \begin{pmatrix}
a_{r} + ia_{i} & b_{r}+ib_{i}\\
-b_{r}+ib_{i} & a_{r}-ia_{i}
\end{pmatrix} \begin{pmatrix}
1 & 0\\
0 & -1
\end{pmatrix}A^{\dagger}\\
%&= \begin{pmatrix}
%a_{r}+ia_{i} & -b_{r}-ib_{i}\\
%-b_{r}+ib_{i} & -a_{r}+ia_{i}
%\end{pmatrix} \begin{pmatrix}
%a_{r}-a_{i} & -b_{r}-ib_{i}\\
%b_{r}-ib_{i} & a_{r}+ia_{i}
%\end{pmatrix}\\
&= \begin{pmatrix}
d_{11} & d_{12}\\
d_{21} & d_{22}
\end{pmatrix}~,
\end{split}
\end{equation}
where
\begin{equation}
\begin{split}
d_{11} &= a_{r}^2 + a_{i}^2 - b_{r}^2 - b_{i}^2~,\\
d_{12} &= -a_{r}b_{r} - ia_{r}b_{i} -ia_{i}b_{r} + a_{i}b_{i} - a_{r}b_{r} - ia_{i}b_{r} - ib_{i}a_{r} - b_{i}a_{i}~,\\
d_{21} &= -b_{r}a_{r} + ib_{r}a_{i}+ib_{i}a_{r}+b_{i}a_{i}-a_{r}b_{r}+ia_{r}b_{i}+ia_{i}b_{r} + a_{i}b_{i}~,\\
d_{22} &= b_{r}^2+b_{i}^2-a_{r}^2-a_{i}^2~.
\end{split}
\end{equation}
As we want
\begin{equation}
A\sigma^{z}A^{\dagger} = \begin{pmatrix}
-1 & 0\\
0 & 1
\end{pmatrix}~,
\end{equation}
we must have
\begin{equation}\label{eq4.51}
\begin{split}
a_{r}^2+a_{i}^2-b_{r}^2-b_{i}^2 &= -1~,\\
b_{i}a_{i} - a_{r}b_{r} &= 0~,\\
b_{i}a_{r} + a_{i}b_{r} &= 0~.
\end{split}
\end{equation}
It is easy to see that a solution is
\begin{equation}
\begin{split}
a_{r} = 0 = a_{i}~,\\
b_{r} = -\frac{1}{\sqrt{2}} = -b_{i}~.
\end{split}
\end{equation}

And so,
\begin{equation}
A = \frac{1}{\sqrt{2}} \begin{pmatrix}
	0 & -1+i\\
	1+i & 0
	\end{pmatrix}=\frac{i}{\sqrt{2}}(\sigma^{x}-\sigma^{y})
	\end{equation}
is the searched matrix.

We note that we have (as expected)
\begin{equation}
\begin{split}
U\sigma^{y}U^{\dagger} &= \frac{1}{2}\begin{pmatrix}
0 & -1+i\\
1+i & 0
\end{pmatrix}\begin{pmatrix}
0 & -i\\
i & 0
\end{pmatrix}\begin{pmatrix}
0 & 1-i\\
-1-i & 0
\end{pmatrix}\\
%&= \frac{1}{2}\begin{pmatrix}
%-i-1 & 0\\
%0 & -i+1
%\end{pmatrix}\begin{pmatrix}
%0 & 1-i\\
%-1-i & 0
%\end{pmatrix}\\
&= \frac{1}{2} \begin{pmatrix}
0 & -2\\
-2 & 0
\end{pmatrix} = -\sigma^{x}~.
\end{split}
\end{equation}

The energy current for the graded $\mathit{XXZ}$ chain is
\begin{equation}
\begin{split}
\hat{J}^{\mathit{XXZ}}= 2\alpha[\alpha(\sigma_{i-1}^{y}\sigma_{i}^{z}\sigma_{i+1}^{x} - \sigma_{i-1}^{x}\sigma_{i}^{z}\sigma_{i+1}^{y})\\ +\Delta_{i-1}(\sigma_{i-1}^{z}\sigma_{i}^{x}\sigma_{i+1}^{y} - \sigma_{i-1}^{z}\sigma_{i}^{y}\sigma_{i+1}^{x})
\\ +\Delta_{i}(\sigma_{i-1}^{x}\sigma_{i}^{y}\sigma_{i+1}^{z}- \sigma_{i-1}^{y}\sigma_{i}^{x}\sigma_{i+1}^{z})]~.
\end{split}
\end{equation}

For the action of $U$, noting that $A^{\dagger}\sigma^{x}A = -\sigma^{y},~~
A^{\dagger}\sigma^{y}A = -\sigma^{x},~~A^{\dagger}\sigma^{z}A = -\sigma^{z}$,
we have
\begin{equation}
\begin{split}
U^{\dagger}\hat{J}^{\mathit{XXZ}}U &= -2\alpha[\alpha(\sigma_{i-1}^{x}\sigma_{i}^{z}\sigma_{i+1}^{y} - \sigma_{i-1}^{y}\sigma_{i}^{z}\sigma_{i+1}^{x})\\&+\Delta_{i-1}(\sigma_{i-1}^{z}\sigma_{i}^{y}\sigma_{i+1}^{x} - \sigma_{i-1}^{z}\sigma_{i}^{x}\sigma_{i+1}^{y})\\&+ \Delta_{i}(\sigma_{i-1}^{y}\sigma_{i}^{x}\sigma_{i+1}^{z} - \sigma_{i-1}^{x}\sigma_{i}^{y}\sigma_{i+1}^{z})]\\
&= 2\alpha[\alpha(\sigma_{i-1}^{y}\sigma_{i}^{z}\sigma_{i+1}^{x} - \sigma_{i-1}^{x}\sigma_{i}^{z}\sigma_{i+1}^{y})\\ &+\Delta_{i-1}(\sigma_{i-1}^{z}\sigma_{i}^{x}\sigma_{i+1}^{y} - \sigma_{i-1}^{z}\sigma_{i}^{y}\sigma_{i+1}^{x})
\\&+\Delta_{i}(\sigma_{i-1}^{x}\sigma_{i}^{y}\sigma_{i+1}^{z}- \sigma_{i-1}^{y}\sigma_{i}^{x}\sigma_{i+1}^{z})]\\
&= \hat{J}^{\mathit{XXZ}}~,
\end{split}
\end{equation}
that is, the one-way street phenomenon holds.

For the spin current
\begin{equation}
\hat{J}^{M} = 2\alpha(\sigma_{i}^{x}\sigma_{i+1}^{y}-\sigma_{i}^{y}\sigma_{i+1}^{x})~,
\end{equation}
it follows
\begin{equation}
\begin{split}
U^{\dagger}\hat{J}^{M}U &= 2\alpha(\sigma_{i}^{y}\sigma_{i+1}^{x} - \sigma_{i}^{x}\sigma_{i+1}^{y})\\ &= -2\alpha(\sigma_{i}^{x}\sigma_{i+1}^{y} - \sigma_{i}^{y}\sigma_{i+1}^{x}) \\ &=-\hat{J}^{M}~,
\end{split}
\end{equation}
that is, the current is inverted without rectification or any other effect.

 {\bf Y-YZ polarization.}
 We now investigate the chain in which the first spin
 is target on $Y$ direction and the last one target on some direction on $YZ$
 plane. We also turn to the Heisenberg models.

  Precisely, we consider the Lindblad operators
\begin{equation}
\begin{split}
K_{+}^{L}&= \sqrt{\gamma (1+f)}\left(\frac{\sigma_{1}^{z}+ i\sigma_{1}^{x}}{2}\right)~, ~~~~
K_{-}^{L}= \sqrt{\gamma (1-f)}\left(\frac{\sigma_{1}^{z}- i\sigma_{1}^{x}}{2}\right)~, \\
K_{+}^{R}&= \sqrt{\gamma (1-f)}\left(\frac{\cos{\theta}\sigma_{N}^{y}+ \sin{\theta}\sigma_{N}^{z}+ i\sigma_{N}^{x}}{2}\right)~,~~~~
K_{-}^{R}= \sqrt{\gamma (1+f)}\left(\frac{\cos{\theta}\sigma_{N}^{y}+ \sin{\theta}\sigma_{N}^{z}- i\sigma_{N}^{x}}{2}\right)~.
\end{split}
\end{equation}

To perform the baths inversion, it is enough to find $A$ such that  $A(\cdot)A^{\dagger}$ makes the transformations
\begin{gather*}
\sigma^{z} \xrightarrow[\text{(1)}]{\text{}} \cos\theta\sigma^{y} + \sin{\theta}\sigma^{z} \xrightarrow[\text{(2)}]{\text{}} \sigma^{z}~,\\
\sigma^{x} \xrightarrow[\text{(3)}]{\text{}} -\sigma^{x}~.
\end{gather*}

We will use the representation of a unitary matrix in $SU(2)$:
\begin{equation}
\begin{split}
A = \begin{pmatrix}
a_{r}+ia_{i} & b_{r}+ib_{i}\\
-b_{r}+ib_{i} & a_{r}-ia_{i}
\end{pmatrix}~,
\end{split}
\end{equation}
where $a_{r}^{2}+a_{i}^{2}+b_{r}^{2}+b_{i}^{2} = 1$

We begin studying the condition $(3)$ above:
\begin{equation}
A\sigma^{x}A^{\dagger} = \begin{pmatrix}
c_{11} & c_{12}\\
c_{21} & c_{22}
\end{pmatrix}~,
\end{equation}
where we want
\begin{equation}
= \begin{pmatrix} 0 & -1 \\
-1 & 0\end{pmatrix}~.
\end{equation}

It leads to
\begin{equation}
\begin{split}
c_{11} &= 2(a_{r}b_{r} + a_{i}b_{i})~, \\
c_{12} &= a_{r}^{2} + b_{i}^{2} - a_{i}^{2} - b_{r}^{2} + 2(a_{i}a_{r} - b_{i}b_{r})i~, \\
c_{21} &= a_{r}^{2} + b_{i}^{2} - a_{i}^{2} - b_{r}^{2} + 2(b_{i}b_{r}-a_{i}a_{r})i~,\\
c_{22} &= -2(a_{r}b_{r}+a_{i}b_{i})~, \\
\end{split}
\end{equation}
and we must have
\begin{equation}
\begin{split}
a_{r}b_{r} &= -a_{i}b_{i}~,\\
a_{r}a_{i} &= b_{r}b_{i}~,\\
a_{r}^{2}+b_{i}^{2}-a_{i}^{2}-b_{i}^{2} &= -1~,\\
a_{r}^{2}+b_{i}^{2}+a_{i}^{2}+b_{r}^{2} &=1~.
\end{split}
\end{equation}

We can take $a_{r} = b_{i} = 0$, and so, we stay with
\begin{equation}
A = \begin{pmatrix}
ia_{i} & b_{r}\\
-b_{r} & -ia_{i}
\end{pmatrix}~,
\end{equation}
where $a_{i}^{2} + b_{r}^{2} =1$.

Let us satisfy condition $(1)$.  We have
\begin{equation}
A\sigma^{z}A^{\dagger} = \begin{pmatrix}
d_{11} & d_{12}\\
d_{21} & d_{22}
\end{pmatrix}~,
\end{equation}
and we want
\begin{equation}
= \begin{pmatrix}
\sin{\theta} & -i\cos{\theta}\\
i\cos{\theta} & -\sin{\theta}
\end{pmatrix}~.
\end{equation}
That is
\begin{equation}
\begin{split}
d_{11}&= a_{i}^{2}-b_{r}^{2}~,\\
d_{12}&= -2ia_{i}b_{r}~,\\
d_{21}&= 2ia_{i}b_{r}~,\\
d_{22}&= b_{r}^{2}-a_{i}^{2}~, \\
\end{split}
\end{equation}
and so
\begin{equation}
\begin{split}
a_{i}^{2} - b_{r}^{2} &= \sin{\theta}~,\\
a_{i}b_{r} = \frac{\cos{\theta}}{2}~.
\end{split}
\end{equation}
Consequently,
\begin{equation}
\begin{split}
a_{i}b_{r} = \frac{\cos{\theta}}{2}=\frac{|\cos{\theta}|}{2}~,
\end{split}
\end{equation}
for $0 \leq \theta \leq \pi/2$ which are our angles of interest.

We take
\begin{equation}
\begin{split}
a_{i} &= \frac{\sqrt{1+\sin{\theta}}}{\sqrt{2}} ~,\\
b_{r} &= \frac{\sqrt{1-\sin{\theta}}}{\sqrt{2}}~.
\end{split}
\end{equation}
 Moreover
\begin{equation}
\begin{split}
a_{i}^{2}+b_{r}^{2} &= \frac{1+\sin{\theta}+1-\sin{\theta}}{2} =1~,\\
a_{i}^{2} - b_{r}^{2} &= \frac{1+\sin{\theta}-1+\sin{\theta}}{2} = \sin{\theta}~.
\end{split}
\end{equation}

Then, the final form of $A$ is
\begin{equation}
A = \frac{1}{\sqrt{2}}\begin{pmatrix}
	i\sqrt{1+\sin{\theta}} & \sqrt{1-\sin{\theta}} \\
	-\sqrt{1-\sin{\theta}} & -i\sqrt{1+\sin{\theta}}
	\end{pmatrix}~.
\end{equation}

We know that
\begin{equation*}
A\sigma^{x}A^{\dagger} = \cos{\theta}\sigma^{y} + \sin{\theta}\sigma^{z}~.
\end{equation*}
A simple computation shows that
\begin{equation*}
A\sigma^{y}A^{\dagger} = \cos{\theta}\sigma^{z}-\sin{\theta}\sigma^{y}~.
\end{equation*}

Noting that $A^{\dagger} = -A$, it follows
\begin{equation*}
A\sigma^{x}A^{\dagger} = (-A)^{\dagger}\sigma^{x}(-A) = A^{\dagger}\sigma^{x}A~,
\end{equation*}
and the same for $\sigma^{y}$ and $\sigma^{z}$.

For the graded Heisenberg Hamiltonian we have
\begin{equation}
\begin{split}
UHU^{\dagger} &= \sum_{i=1}^{N-1}\alpha_{i}\{(-\sigma_{i}^{x})(-\sigma_{i+1}^{x}) +(\cos{\theta}\sigma_{i}^{z}-\sin{\theta}\sigma_{i}^{y})(\cos{\theta}\sigma_{i+1}^{z}-\sin{\theta}\sigma_{i+1}^{y}) \\&+(\cos{\theta}\sigma_{i}^{y} + \sin{\theta}\sigma_{i}^{z})(\cos{\theta}\sigma_{i+1}^{y} + \sin{\theta}\sigma_{i+1}^{z})\}\\
&= \sum_{i=1}^{N-1}\alpha_{i}\{\sigma_{i}^{x}\sigma_{i}^{y} + \cos^{2}{\theta}\sigma_{i}^{z}\sigma_{i+1}^{z} - \cos{\theta}\sin{\theta}\sigma_{i}^{z}\sigma_{i+1}^{y}\\&- \sin{\theta}\cos{\theta}\sigma_{i}^{y}\sigma_{i+1}^{z} + \sin^{2}{\theta}\sigma_{i}^{y}\sigma_{i+1}^{y} + \cos^{2}{\theta}\sigma_{i}^{y}\sigma_{i+1}^{y}\\&+ \cos{\theta}\sin{\theta}\sigma_{i}^{y}\sigma_{i+1}^{z} + \sin{\theta}\cos{\theta}\sigma_{i}^{z}\sigma_{i}^{y} + \sin^{2}{\theta}\sigma_{i}^{z}\sigma_{i+1}^{z} \}\\
&= \sum_{i=1}^{N-1}\alpha_{i}\{\sigma_{i}^{x}\sigma_{i+1}^{x} + \sigma_{i}^{z}\sigma_{i+1}^{z} + \sigma_{i}^{y}\sigma_{i+1}^{y}\} = H~.
\end{split}
\end{equation}

For the energy current
\begin{equation}
\begin{split}
U\hat{J}^{E}U^{\dagger} &= U\{(\sigma_{i-1}^{x}\sigma_{i}^{y}\sigma_{i+1}^{z}+ \sigma_{i-1}^{y}\sigma_{i}^{z}\sigma_{i+1}^{x} \\&+ \sigma_{i-1}^{z}\sigma_{i}^{x}\sigma_{i+1}^{y} - \sigma_{i-1}^{x}\sigma_{i}^{z}\sigma_{i+1}^{y} \\&-\sigma_{i-1}^{y}\sigma_{i}^{x}\sigma_{i+1}^{z} -\sigma_{i-1}^{z}\sigma_{i}^{y}\sigma_{i+1}^{x})\}U^{\dagger}~,
\end{split}
\end{equation}
we have
\begin{equation}
\begin{split}
U\hat{J}^{E}U^{\dagger} &= 2\alpha_{i-1}\alpha_{i}\{(-\sigma_{i-1}^{x})(\cos{\theta}\sigma_{i}^{z} - \sin{\theta}\sigma_{i}^{y})(\cos{\theta}\sigma_{i+1}^{y}+\sin{\theta}\sigma_{i+1}^{z}) \\&+ (\cos{\theta}\sigma_{i-1}^{z} - \sin{\theta}\sigma_{i-1}^{y})(\cos{\theta}\sigma_{i}^{y}+\sin{\theta}\sigma_{i}^{z})(-\sigma_{i+1}^{x}) \\&+(\cos{\theta}\sigma_{i-1}^{y}+\sin{\theta}\sigma_{i}^{z})(-\sigma_{i}^{x})(\cos{\theta}\sigma_{i+1}^{z}-\sin{\theta} \sigma_{i+1}^{y})
\\&-(-\sigma_{i-1}^{x})(\cos{\theta}\sigma_{i}^{y} + \sin{\theta}\sigma_{i}^{z})(\cos{\theta}\sigma_{i+1}^{z} - \sin{\theta}\sigma_{i+1}^{y})
\\&-(\cos{\theta}\sigma_{i-1}^{z} - \sin{\theta}\sigma_{i-1}^{y})(-\sigma_{i}^{x})(\cos{\theta}\sigma_{i+1}^{y} + \sin{\theta}\sigma_{i+1}^{z})\\
&-(\cos{\theta}\sigma_{i-1}^{y} + \sin{\theta}\sigma_{i-1}^{z})(\cos{\theta}\sigma_{i}^{z} - \sin{\theta}\sigma_{i}^{y})(-\sigma_{i+1}^{x}) \}~,
\end{split}
\end{equation}
and so
\begin{equation}
\begin{split}
U\hat{J}^{E}U^{\dagger}&= 2\alpha_{i-1}\alpha_{i}\{-\cos^{2}{\theta}\sigma_{i-1}^{x}\sigma_{i}^{z}\sigma_{i+1}^{y} - \cos{\theta}\sin{\theta}\sigma_{i+1}^{x}\sigma_{i}^{z}\sigma_{i+1}^{z}
\\&+\sin{\theta}\cos{\theta}\sigma_{i-1}^{x}\sigma_{i}^{y}\sigma_{i+1}^{y} + \sin^{2}{\theta}\sigma_{i-1}^{x}\sigma_{i}^{y}\sigma_{i+1}^{z}
\\&-\cos^{2}{\theta}\sigma_{i-1}^{z}\sigma_{i}^{y}\sigma_{i+1}^{x} - \cos{\theta}\sin{\theta}\sigma_{i-1}^{z}\sigma_{i}^{z}\sigma_{i+1}^{x}
\\&+\sin{\theta}\cos{\theta}\sigma_{i-1}^{y}\sigma_{i}^{y}\sigma_{i+1}^{x} + \sin^{2}{\theta}\sigma_{i-1}^{y}\sigma_{i}^{z}\sigma_{i+1}^{x}
\\&-\cos^{2}{\theta}\sigma_{i-1}^{y}\sigma_{i}^{x}\sigma_{i+1}^{z} - \sin{\theta}\cos{\theta}\sigma_{i-1}^{z}\sigma_{i}^{x}\sigma_{i+1}^{z}
\\&+\cos{\theta}\sin{\theta}\sigma_{i-1}^{y}\sigma_{i}^{x}\sigma_{i+1}^{y} + \sin^{2}{\theta}\sigma_{i-1}^{z}\sigma_{i}^{x}\sigma_{i+1}^{y}
\\&+\cos^{2}{\theta}\sigma_{i-1}^{x}\sigma_{i}^{y}\sigma_{i+1}^{z} - \cos{\theta}\sin{\theta}\sigma_{i-1}^{x}\sigma_{i}^{y}\sigma_{i+1}^{y}
\\&+ \sin{\theta}\cos{\theta}\sigma_{i-1}^{x}\sigma_{i}^{z}\sigma_{i+1}^{z} - \sin^{2}{\theta}\sigma_{i-1}^{x}\sigma_{i}^{z}\sigma_{i+1}^{y}
\\&+ \cos^{2}{\theta}\sigma_{i-1}^{z}\sigma_{i}^{x}\sigma_{i+1}^{y} + \cos{\theta}\sin{\theta}\sigma_{i-1}^{z}\sigma_{i}^{x}\sigma_{i+1}^{z}
\\&- \sin{\theta}\cos{\theta}\sigma_{i-1}^{y}\sigma_{i}^{x}\sigma_{i+1}^{y} - \sin^{2}{\theta}\sigma_{i-1}^{y}\sigma_{i}^{x}\sigma_{i+1}^{z}
\\&+\cos^{2}{\theta}\sigma_{i-1}^{y}\sigma_{i}^{z}\sigma_{i+1}^{x} - \cos{\theta}\sin{\theta}\sigma_{i-1}^{y}\sigma_{i}^{y}\sigma_{i+1}^{x}
\\&+ \sin{\theta}\cos{\theta}\sigma_{i-1}^{y}\sigma_{i}^{z}\sigma_{i+1}^{x} - \sin^{2}{\theta}\sigma_{i-1}^{z}\sigma_{i}^{y}\sigma_{i+1}^{x}\}
\\&= 2\alpha_{i-1}\alpha_{i}\{-\sigma_{i-1}^{x}\sigma_{i}^{z}\sigma_{i+1}^{y} + \sigma_{i-1}^{x}\sigma_{i}^{y}\sigma_{i+1}^{z} \\&-\sigma_{i-1}^{z}\sigma_{i}^{y}\sigma_{i+1}^{x} +\sigma_{i-1}^{y}\sigma_{i}^{z}\sigma_{i+1}^{x}
\\&-\sigma_{i-1}^{y}\sigma_{i}^{x}\sigma_{i+1}^{z}+\sigma_{i-1}^{z}\sigma_{i}^{x}\sigma_{i+1}^{y} \}
\\&= \hat{J}^{E}~,
\end{split}
\end{equation}
that shows the occurrence of the one-way street phenomenon.

{\bf Y-Z orthogonal polarizations.}
Let us consider the set of Lindblad
operators at one boundary, say the left one, given by
\begin{equation}
\begin{split}
L_{1} &= \alpha(\sigma_{1}^{x} + i\sigma_{1}^{y})~, ~~~~
L_{2} = \beta(\sigma_{1}^{x} - i\sigma_{1}^{y})~,~~~~
V_{1} = p(\sigma_{1}^{y} + i\sigma_{1}^{z})~,\\
V_{2} &= q(\sigma_{1}^{y} - i\sigma_{1}^{z})~,~~~~
W_{1} = u(\sigma_{1}^{z} + i\sigma_{1}^{x})~,~~~~
W_{2} = v(\sigma_{1}^{z} - i\sigma_{1}^{x})~,
\end{split}
\end{equation}
and for the right boundary
\begin{equation}
\begin{split}
L_{3} &= v(\sigma_{N}^{x} + i\sigma_{N}^{y})~,~~~~
L_{4} = u(\sigma_{N}^{x} - i\sigma_{N}^{y})~,~~~~
V_{3} = q(\sigma_{N}^{y} + i\sigma_{N}^{z})~,\\
V_{4} &= p(\sigma_{N}^{y} - i\sigma_{N}^{z})~,~~~~
W_{3} = \beta(\sigma_{N}^{z} + i\sigma_{N}^{x})~,~~~~
W_{4} = \alpha(\sigma_{N}^{z} - i\sigma_{N}^{x})~.
\\    \end{split}
\end{equation}

To invert the baths it is enough to find an operator $A$ such that
\begin{equation}
\begin{split}
(I)A\sigma^{x}A^{\dagger} = -\sigma^{x} &\Leftrightarrow -\sigma^{x} = A^{\dagger}\sigma^{x}A~,\\
(II)A\sigma^{y}A^{\dagger} = -\sigma^{z} &\Leftrightarrow -\sigma^{y} = A^{\dagger}\sigma^{z}A~,\\
(III)A\sigma^{z}A^{\dagger} = -\sigma^{y} &\Leftrightarrow -\sigma^{z} = A^{\dagger}\sigma^{y}A~.
\end{split}
\end{equation}
Indeed, in such case, we will have the transformations
\begin{equation}
\begin{split}
L_{1} &\rightarrow -iW_{4}~,~~~~
L_{2} \rightarrow iW_{3}~,~~~~
V_{1} \rightarrow -iV_{4}~,\\
V_{2} &\rightarrow iV_{3}~,~~~~
W_{1} \rightarrow -iL_{4}~,~~~~
W_{2} \rightarrow iL_{3}~,
\end{split}
\end{equation}
and also
\begin{equation}
\begin{split}
L_{3} &\rightarrow -iW_{2}~,~~~~
L_{4} \rightarrow iW_{1}~,~~~~
V_{3} \rightarrow -iV_{3}~,\\
V_{4} &\rightarrow iV_{i}~,~~~~
W_{3} \rightarrow -iL_{2}~,~~~~
W_{2} \rightarrow iL_{i}~.
\end{split}
\end{equation}
Consequently, the dissipator transforms as
\begin{equation}
U\mathcal{L}(\rho)U^{\dagger} = \mathcal{L}(\rho, {\rm inverted baths})~.
\end{equation}

As we show below, it is enough to use a representation for $A$ in $SU(2)$
\begin{equation*}
A = \begin{pmatrix}
a_{r}+ ia_{i} & b_{r}+ib_{i}\\
-b_{r} + ib_{i} & a_{r}-ia_{i}
\end{pmatrix}~,
\end{equation*}
where $a_{r},a_{i},b_{r}$ e $b_{i} \in \mathbb{R}$ e $a_{r}^2 + a_{i}^2 + b_{r}^2 + b_{i}^2 = 1$.

Computing $(I)$:
\begin{equation}
\begin{split}
A\sigma^{x}A^{\dagger}
&= \begin{pmatrix}
z_{11} & z_{12}\\
z_{21} & z_{22}
\end{pmatrix}~,
\end{split}
\end{equation}
where
\begin{equation}
\begin{split}
z_{11} &= b_{r}a_{r} -ib_{r}a_{i} +ib_{i}a_{r} + b_{i}a_{i} + b_{r}a_{r} -ia_{r}b_{i}+ib_{r}a_{i}+a_{i}b_{i}~,\\
z_{12} &= -b_{r}^2-ib_{i}b_{r}-ib_{i}b_{r}+b_{i}^2+a_{r}^2+ia_{i}a_{r} + ia_{i}a_{r} - a_{i}^2~,\\
z_{21} &= a_{r}^2 -ia_{r}a_{i} -ia_{i}a_{r}-a_{i}^2-b_{r}^2+ib_{i}b_{r}+ib_{i}b_{r}+b_{i}^2~,\\
z_{22} &= -a_{r}b_{r} -ib_{i}a_{r} + ib_{r}a_{i} -a_{i}b_{i} - b_{r}a_{r} - ia_{i}b_{r}+ib_{i}a_{r} -a_{i}b_{i}~.
\end{split}
\end{equation}
We want
\begin{equation*}
A\sigma^{x}A^{\dagger} =\begin{pmatrix}
0 & -1 \\
-1 & 0
\end{pmatrix}~.
\end{equation*}
And so,
\begin{equation*}\label{eq4.31}
a_{i}b_{i} + a_{r}b_{r} = 0~,
\end{equation*}
\begin{equation*}\label{eq4.32}
a_{r}^2 - a_{i}^2 + b_{i}^2 - b_{r}^2 = -1~,
\end{equation*}
\begin{equation*}\label{eq4.33}
b_{i}b_{r} = a_{i}a_{r}~.
\end{equation*}

We still want from $(II)$
\begin{equation}
A\sigma^{y}A^{\dagger} = \begin{pmatrix}
-1 & 0\\
0 & 1
\end{pmatrix}~,
\end{equation}
that leads us to
\begin{equation*} \label{eq4.34}
a_{i}b_{r} - a_{r}b_{i} = -\frac{1}{2}~,
\end{equation*}
\begin{equation*} \label{eq4.35}
a_{i}a_{r} = -b_{i}b_{r}~,
\end{equation*}
\begin{equation*} \label{eq4.36}
a_{r}^2 - a_{i}^2 + b_{r}^2 - b_{i}^2 = 0~.
\end{equation*}

From the equations above we have $a_{i}a_{r} = 0 = b_{i}b_{r}$. We choose $a_{r} = b_{i} =0$ and, consequently,
\begin{equation*}
a_{i}^{2} = b_{r}^2 = -(-1+b_{r}^2) \Rightarrow -2a_{i}^2 = -1 \Rightarrow a_{i} = \frac{1}{\sqrt{2}}~.
\end{equation*}
Hence,
\begin{equation}
b_{r} = -\frac{1}{\sqrt{2}}~.
\end{equation}

And we obtain for $A$ the final form
\begin{equation}
A =\frac{1}{\sqrt{2}}\begin{pmatrix}
	i & -1\\
	1 & -i
	\end{pmatrix} = \frac{i}{\sqrt{2}}(-\sigma^{z}+\sigma^{y})~.
\end{equation}

A short computation shows us that $(III)$ follows:
\begin{equation*}
A\sigma^{z}A^{\dagger} = -\sigma^{y}~,
\end{equation*}
as expected.

It is easy to see that the transformations keep the Hamiltonian of the graded
Heisenberg model unchanged, i.e.,
\begin{equation}
UHU^{\dagger} = \frac{1}{2}\sum_{i=1}^{N-1}\alpha_{i} (\sigma_{i}^{x}\sigma_{i+1}^{x} + \sigma_{i}^{z}\sigma_{i+1}^{z} +\sigma_{i}^{y}\sigma_{i+1}^{y}) = H~.
\end{equation}

For the energy current, we have
\begin{equation}
\begin{split}
U^{\dagger}\hat{J}^{E}U &= 2\alpha_{i-1}\alpha_{i}(-\sigma_{i-1}^{x}\sigma_{i}^{z}\sigma_{i+1}^{y}-\sigma_{i-1}^{z}\sigma_{i}^{y}\sigma_{i+1}^{x}\\&-\sigma_{i-1}^{y}\sigma_{i}^{x}\sigma_{i+1}^{z}+\sigma_{i-1}^{x}\sigma_{i}^{y}\sigma_{i+1}^{z}\\&+\sigma_{i-1}^{z}\sigma_{i}^{x}\sigma_{i+1}^{y} + \sigma_{i-1}^{y}\sigma_{i}^{z}\sigma_{i+1}^{x})\\ &= \hat{J}^{E}~,
\end{split}
\end{equation}
that is, the one-way street phenomenon holds.

{\bf Z-XZ polarization.}
 We now consider the case involving a $\sigma^{z}$ target polarization at one side, and on a rotated axis on plane $XZ$ for the other side. That is, we consider the Lindblad operators as
\begin{equation*}
\begin{split}
K_{+}^{L} &= \sqrt{\gamma(1+f)}\left( \frac{\sigma_{1}^{x} + i\sigma_{1}^{y}}{2}\right)~,~~~~
K_{-}^{L} = \sqrt{\gamma(1-f)}\left(\frac{\sigma_{1}^{x} - i\sigma_{1}^{y}}{2}\right)~,\\
K_{+}^{R} &= \sqrt{\gamma(1-f)}\left(\frac{\cos{\theta}\sigma_{N}^{x} + \sin{\theta}\sigma_{N}^{z} + i\sigma^{y}}{2}\right)~,~~~~
K_{-}^{R} = \sqrt{\gamma(1+f)}\left(\frac{\cos{\theta}\sigma_{N}^{x} + \sin{\theta}\sigma_{N}^{z}-i\sigma^{y}}{2}\right)~.
\end{split}
\end{equation*}

Again, we search for one operator related to baths inversion. We use the general representation of $SU(2)$. After manipulations similar to those previously described, we find
\begin{equation}
A=\frac{i}{\sqrt{2}}\begin{pmatrix}
	\sqrt{1-\cos{\theta}} & \sqrt{1+\cos{\theta}} \\
	\sqrt{1+\cos{\theta}} & -\sqrt{1-\cos{\theta}}
	\end{pmatrix}~.
\end{equation}

Then we study the effect of $U = A \otimes A \otimes ... \otimes A$ on the Heisenberg Hamiltonian and on the energy current.
We have
\begin{equation}
\begin{split}
UHU^{\dagger} &= \ldots\\
&=\frac{1}{2}\sum_{i=1}^{N-1}\alpha_{i}[(\cos{\theta}\sigma_{i}^{x}+\sin{\theta}\sigma_{i}^{z})(\cos{\theta}\sigma_{i+1}^{x} + \sin{\theta}\sigma_{i+1}^{z})\\
&+ (-\sigma_{i}^{y})(-\sigma_{i+1}^{y})+(\sin{\theta}\sigma_{i}^{x} - \cos{\theta}\sigma_{i}^{z})(\sin{\theta}\sigma_{i+1}^{x}-\cos{\theta}\sigma_{i+1}^{z})]\\
&=\frac{1}{2}\sum_{i=1}^{N-1}\alpha_{i}[\cos^2{\theta}\sigma_{i}^{x}\sigma_{i+1}^{x} + \cos{\theta}\sin{\theta}\sigma_{i}^{x}\sigma_{i+1}^{z} \\ &+ \sin{\theta}\cos{\theta}\sigma_{i}^{z}\sigma_{i+1}^{x} + \sin^2{\theta}\sigma_{i}^{z}\sigma_{i+1}^{z} \\ &+ \sigma_{i}^{y}\sigma_{i+1}^{y} + \sin^2{\theta}\sigma_{i}^{x}\sigma_{i+1}^{x} - \sin{\theta}\cos{\theta}\sigma_{i}^{x}\sigma_{i+1}^{z}\\ &-\cos{\theta}\sin{\theta}\sigma_{i}^{z}\sigma_{i+1}^{x} + \cos^2{\theta}\sigma_{i}^{z}\sigma_{i+1}^{z}]\\
&=\frac{1}{2}\sum_{i=1}^{N-1}\alpha_{i} (\sigma_{i}^{x}\sigma_{i+1}^{x} + \sigma_{i}^{y}\sigma_{i+1}^{y} +\sigma_{i}^{z}\sigma_{i+1}^{z}) \\
&= H~,
\end{split}
\end{equation}
as expected.

For the energy current of the Heisenberg model we have
\begin{equation}
\begin{split}
U^{\dagger}\hat{J}^{E}U &= 2\alpha_{i-1}\alpha_{i}[(-\sigma_{i-1}^{y})(\sin{\theta}\sigma_{i}^{x}-\cos{\theta}\sigma_{i}^{z})(\cos{\theta}\sigma_{i+1}^{x}+\sin{\theta}\sigma_{i+1}^{z})\\&-(\cos{\theta}\sigma_{i-1}^{x}+\sin{\theta}\sigma_{i-1}^{z})(\sin{\theta}\sigma_{i}^{x}-\cos{\theta}\sigma_{i}^{z})(-\sigma_{i+1}^{y})\\&+(\sin{\theta}\sigma_{i-1}^{x}-\cos{\theta}\sigma_{i-1}^{z})(\cos{\theta}\sigma_{i}^{x}+\sin{\theta}\sigma_{i}^{z})(-\sigma_{i+1}^{y})\\&-(-\sigma_{i-1}^{y})(\cos{\theta}\sigma_{i}^{x}+\sin{\theta}\sigma_{i}^{z})(\sin{\theta}\sigma_{i+1}^{x}-\cos{\theta}\sigma_{i+1}^{z})\\&+(\cos{\theta}\sigma_{i-1}^{x}+\sin{\theta}\sigma_{i-1}^{z})(-\sigma_{i}^{y})(\sin{\theta}\sigma_{i+1}^{x}-\cos{\theta}\sigma_{i+1}^{z})\\&-(\sin{\theta}\sigma_{i-1}^{x} -\cos{\theta}\sigma_{i-1}^{z})(-\sigma_{i}^{y})(\cos{\theta}\sigma_{i+1}^{x}+\sin{\theta}\sigma_{i+1}^{z})]\\
&=2\alpha_{i-1}\alpha_{i}[(-\sin{\theta}\sigma_{i-1}^{y}\sigma_{i}^{x}+\cos{\theta}\sigma_{i-1}^{y}\sigma_{i}^{z})(\cos{\theta}\sigma_{i+1}^{x}+\sin{\theta}\sigma_{i+1}^{z})\\&-(\cos{\theta}\sigma_{i-1}^{x}+\sin{\theta}\sigma_{i-1}^{z})(-\sin{\theta}\sigma_{i}^{x}\sigma_{i+1}^{y}+\cos{\theta}\sigma_{i}^{z}\sigma_{i+1}^{y})\\&+(\sin{\theta}\sigma_{i-1}^{x}+\cos{\theta}\sigma_{i-1}^{z})(-\cos{\theta}\sigma_{i}^{x}\sigma_{i+1}^{y}-\sin{\theta}\sigma_{i}^{z}\sigma_{i+1}^{y})\\&+(\cos{\theta}\sigma_{i-1}^{y}\sigma_{i}^{x}+\sin{\theta}\sigma_{i-1}^{y}\sigma_{i}^{z})(\sin{\theta}\sigma_{i+1}^{x}-\cos{\theta}\sigma_{i+1}^{z})\\&+(\cos{\theta}\sigma_{i-1}^{x}+\sin{\theta}\sigma_{i-1}^{z})(-\sin{\theta}\sigma_{i}^{y}\sigma_{i+1}^{x}+\cos{\theta}\sigma_{i}^{y}\sigma_{i+1}^{z})\\&+(\sin{\theta}\sigma_{i-1}^{x}-\cos{\theta}\sigma_{i-1}^{z})(\cos{\theta}\sigma_{i}^{y}\sigma_{i+1}^{x} + \sin{\theta}\sigma_{i}^{y}\sigma_{i+1}^{z})] \\
&=2\alpha_{i-1}\alpha_{i}[-\sin{\theta}\cos{\theta}\sigma_{i-1}^{y}\sigma_{i}^{x}\sigma_{i+1}^{x} -\sin^2{\theta}\sigma_{i-1}^{y}\sigma_{i}^{x}\sigma_{i+1}^{z} \\&+ \cos^2{\theta}\sigma_{i-1}^{y}\sigma_{i}^{z}\sigma_{i+1}^{x} + \sin{\theta}\cos{\theta}\sigma_{i-1}^{y}\sigma_{i}^{z}\sigma_{i+1}^{z}\\ &+ \cos{\theta}\sin{\theta}\sigma_{i-1}^{x}\sigma_{i}^{x}\sigma_{i+1}^{y} -\cos^2{\theta}\sigma_{i-1}^{x}\sigma_{i}^{z}\sigma_{i+1}^{y}\\ &+ \sin^2{\theta}\sigma_{i-1}^{z}\sigma_{i}^{x}\sigma_{i+1}^{y} - \sin{\theta}\cos{\theta}\sigma_{i-1}^{z}\sigma_{i}^{z}\sigma_{i+1}^{y} \\ &-\sin{\theta}\cos{\theta}\sigma_{i-1}^{x}\sigma_{i}^{x}\sigma_{i+1}^{y} - \sin^2{\theta}\sigma_{i-1}^{x}\sigma_{i}^{z}\sigma_{i+1}^{y} \\&+ \cos^2{\theta}\sigma_{i-1}^{z}\sigma_{i}^{x}\sigma_{i+ 1}^{y} + \cos{\theta}\sin{\theta}\sigma_{i-1}^{z}\sigma_{i}^{z}\sigma_{i+1}^{y}\\&+\cos{\theta}\sin{\theta}\sigma_{i-1}^{y}\sigma_{i}^{x}\sigma_{i+1}^{x}-\cos^2{\theta}\sigma_{i-1}^{y}\sigma_{i}^{x}\sigma_{i+1}^{z}\\&+\sin^2{\theta}\sigma_{i-1}^{y}\sigma_{i}^{z}\sigma_{i+1}^{x}-\sin{\theta}\cos{\theta}\sigma_{i-1}^{y}\sigma_{i}^{z}\sigma_{i+1}^{z} \\&-\cos{\theta}\sin{\theta}\sigma_{i-1}^{x}\sigma_{i}^{y}\sigma_{i+1}^{x} + \cos^2{\theta}\sigma_{i-1}^{x}\sigma_{i}^{y}\sigma_{i+1}^{z}\\&-\sin^2{\theta}\sigma_{i-1}^{z}\sigma_{i}^{y}\sigma_{i+1}^{x} + \sin{\theta}\cos{\theta}\sigma_{i-1}^{z}\sigma_{i}^{y}\sigma_{i+1}^{z} \\ &+\sin{\theta}\cos{\theta}\sigma_{i-1}^{x}\sigma_{i}^{y}\sigma_{i+1}^{x}+\sin^2{\theta}\sigma_{i-1}^{x}\sigma_{i}^{y}\sigma_{i+1}^{z} \\&- \cos^2{\theta}\sigma_{i-1}^{z}\sigma_{i}^{y}\sigma_{i+1}^{x} -\cos{\theta}\sin{\theta}\sigma_{i-1}^{z}\sigma_{i}^{y}\sigma_{i+1}^{z}]~,
\end{split}
\end{equation}
and using $\cos^2{x}+\sin^2{x}=1$, we obtain
\begin{equation}
\begin{split}
U^{\dagger}\hat{J}^{E}U &= 2\alpha_{i-1}\alpha_{i}[-\sigma_{i-1}^{y}\sigma_{i}^{x}\sigma_{i+1}^{z} + \sigma_{i-1}^{y}\sigma_{i}^{z}\sigma_{i+1}^{x} \\&- \sigma_{i-1}^{x}\sigma_{i}^{z}\sigma_{i+1}^{y} + \sigma_{i-1}^{z}\sigma_{i}^{x}\sigma_{i+1}^{y} \\&+ \sigma_{i-1}^{x}\sigma_{i}^{y}\sigma_{i+1}^{z} - \sigma_{i-1}^{z}\sigma_{i}^{y}\sigma_{i+1}^{x}]\\
&= 2\alpha_{i-1}\alpha_{i}[ \sigma_{i-1}^{y}\sigma_{i}^{z}\sigma_{i+1}^{x} - \sigma_{i-1}^{x}\sigma_{i}^{z}\sigma_{i+1}^{y}\\&+\sigma_{i-1}^{z}\sigma_{i}^{x}\sigma_{i+1}^{y} - \sigma_{y}^{i-1}\sigma_{i}^{x}\sigma_{i+1}^{z}\\ &+\sigma_{i-1}^{x}\sigma_{i}^{y}\sigma_{i+1}^{z} - \sigma_{i-1}^{z}\sigma_{i}^{y}\sigma_{i+1}^{x}] = \hat{J}^{E}~,
\end{split}
\end{equation}
that is, one-way street phenomenon.

{\bf X-Z orthogonal polarization.} Now, for one boundary we take the Lindblad operators
\begin{equation*}
\begin{split}
L_{1} &= \alpha(\sigma_{1}^{x} + i\sigma_{1}^{y})~,~~~~
L_{2} = \beta(\sigma_{1}^{x} - i\sigma_{1}^{y})~,\\
V_{1} &= p(\sigma_{1}^{y} + i\sigma_{1}^{z})~,~~~~
V_{2} = q(\sigma_{1}^{y} - i\sigma_{1}^{z})~,\\
W_{1} &= u(\sigma_{1}^{z} + i\sigma_{1}^{x})~,~~~~
W_{2} = v(\sigma_{1}^{z} - i\sigma_{1}^{x})~,
\end{split}
\end{equation*}
and, for the opposite boundary,
\begin{equation}
\begin{split}
L_{3} &= q(\sigma_{N}^{x} + i\sigma_{N}^{y})~,~~~~
L_{4} = p(\sigma_{N}^{x} - i\sigma_{N}^{y})~,\\
V_{3} &= \beta(\sigma_{N}^{y} + i\sigma_{N}^{z})~,~~~~
V_{4} = \alpha(\sigma_{N}^{y} - i\sigma_{N}^{z})~,\\
W_{3} &= v(\sigma_{N}^{z} + i\sigma_{N}^{x})~,~~~~
W_{4} = u(\sigma_{N}^{z} - i\sigma_{N}^{x})~.
\\    \end{split}
\end{equation}

To implement the bath inversion, it is enough to find a unitary operator $A$ such that
\begin{equation}
\begin{split}
(I)A\sigma^{y}A^{\dagger} = -\sigma^{y} &\Leftrightarrow -\sigma^{y} = A^{\dagger}\sigma^{y}A~,\\
(II)A\sigma^{x}A^{\dagger} = -\sigma^{z} &\Leftrightarrow -\sigma^{x} = A^{\dagger}\sigma^{z}A~,\\
(III)A\sigma^{z}A^{\dagger} = -\sigma^{x} &\Leftrightarrow -\sigma^{z} = A^{\dagger}\sigma^{x}A~,
\end{split}
\end{equation}
since its action will perform the transformations
\begin{equation}
\begin{split}
L_{1} &\rightarrow -iV_{4}~,~~~~ L_{2} \rightarrow iV_{3}~,\\
V_{1} &\rightarrow -iL_{4}~,~~~~ V_{2} \rightarrow iL_{3}~,\\
W_{1} &\rightarrow -iW_{4}~,~~~~ W_{2} \rightarrow iW_{3}~,
\end{split}
\end{equation}
and also
\begin{equation}
\begin{split}
L_{3} &\rightarrow -iV_{2}~,~~~~
L_{4} \rightarrow iV_{1}~,\\
V_{3} &\rightarrow -iL_{3}~,~~~~
V_{4} \rightarrow iL_{i}~,\\
W_{3} &\rightarrow -iW_{2}~,~~~~
W_{2} \rightarrow iW_{i}~.
\end{split}
\end{equation}

After some algebraic manipulations, we find
\begin{equation}
A = \frac{1}{\sqrt{2}}\begin{pmatrix}
	-i & i\\
	i & i
	\end{pmatrix} = \frac{i}{\sqrt{2}}(\sigma^{x}-\sigma^{z})~.
\end{equation}

And everything follows: the Heisenberg Hamiltonian is preserved under the transformations, as well as the energy current, i.e., the one-way street phenomenon holds.

\section{Steady state uniqueness}

Now we prove the uniqueness of the steady state for all the cases previously
analyzed here.

As well known, the steady state is unique if the set of Lindblad operators
together with the Hamiltonian are enough to generate the whole Pauli algebra
\cite{Prosen2} involving all sites $1, 2, \ldots, N$. Here, in our
prove, we follow Prosen \cite{Prosen2}.

In any of the previous analyzed cases, the Lindblad operators are given in terms of 
 $\sigma^{+}$ and $\sigma^{-}$, or $\Gamma^{+}=\frac{\sigma^{z} + i\sigma^{x}}{2}$ and $\Gamma^{-}=\frac{\sigma^{z} - i\sigma^{x}}{2}$
 in one of the sides of the system ($1$ or $N$), or in terms of  $\Pi^{+} = \frac{\sigma^{y}+i\sigma^{z}}{2}$ and $\Pi^{-} = \frac{\sigma^{y}-i\sigma^{z}}{2}$ . But this last case is reduced to the first one by the relations
\begin{equation}
\begin{split}
[\Pi^{+},\Pi^{-}] = \sigma^{x}~,\\
\Pi^{+} + \Pi^{-} = \sigma^{y}~,\\
(\Pi^{+} - \Pi^{-})(-i) = \sigma^{z}~,
\end{split}
\end{equation}
and the other reduces to the first one due
\begin{equation}
\begin{split}
-i[\Gamma^{+},\Gamma^{-}] = \sigma^{y}~,\\
\Gamma^{+} + \Gamma^{-} = \sigma^{z}~,\\
(\Gamma^{+} - \Gamma^{-})(-i) = \sigma^{x}~.
\end{split}
\end{equation}

Thus, let us show that having
 $\sigma^{+}$ and $\sigma^{-}$ in one of the sides is enough to generate the
 whole algebra (of course, with the Hamiltonian). To prove it, we will show
 that the following relations are valid
\begin{equation}
\begin{split}
\sigma_{2}^{+} &= \frac{1}{4}\sigma_{1}^{z}[\sigma_{1}^{+},[H,\sigma_{1}^{z}]]~,\\
\sigma_{j}^{+} &= -\sigma_{j-2}^{+}-\frac{1}{2}\sigma_{j-1}^{z}[\sigma_{j-1}^{-},\sigma_{j-1}^{+}H\sigma_{j-1}^{+}]~,
\end{split}
\end{equation}
for $j=3,4,...,n$ and the conjugate
\begin{equation}
\begin{split}
\sigma_{2}^{-} &= \frac{1}{4}[\sigma_{1}^{-},[H,\sigma_{1}^{z}]]\sigma_{1}^{z}~,\\
\sigma_{j}^{-} &= -\sigma_{j-2}^{-} + \frac{1}{2}[\sigma_{j-1}^{+},\sigma_{j-1}^{-}H\sigma_{j-1}^{-}]\sigma_{j-1}^{z}~.
\end{split}
\end{equation}
we recall that $[\sigma^{+},\sigma^{-}]=\sigma^{z}$. With the previous relations, we get the set $\{\sigma_{j}^{+},\sigma_{j}^{-};j=1,...,n\}$ that
generates the whole Pauli algebra.

First, we rewrite the $\mathit{XXZ}$ Hamiltonian as
\begin{equation}
H = \sum_{j=1}^{n-1}(2\sigma_{j}^{+}\sigma_{j+1}^{-} + 2\sigma_{j}^{-}\sigma_{j+1}^{+} + \Delta\sigma_{j}^{z}\sigma_{j+1}^{z})~.
\end{equation}
Talking about algebraic properties, the constants $\alpha$ and $\Delta$ are
not important (as well as the difference between $\Delta_{j}$ and $\Delta_{j+1}$). And so, our computation follows also for the Heisenberg model.

We have
\begin{equation}
\begin{split}
[H,\sigma_{1}^{z}] &= 2[\sigma_{1}^{+}\sigma_{2}^{-},\sigma_{1}^{z}] + 2[\sigma_{1}^{-}\sigma_{2}^{+},\sigma_{1}^{z}]\\
&=2[\sigma_{1}^{+},\sigma_{1}^{z}]\sigma_{2}^{-} + 2[\sigma_{1}^{-},\sigma_{1}^{z}]\sigma_{2}^{+} \\&= -4\sigma_{1}^{+}\sigma_{2}^{-} + 4\sigma_{1}^{-}\sigma_{2}^{+}~,
\end{split}
\end{equation}
and so
\begin{equation}
\begin{split}
[\sigma_{1}^{+},[H,\sigma_{1}^{z}]] &= -4[\sigma_{1}^{+},\sigma_{1}^{+}\sigma_{2}^{-}] + 4[\sigma_{1}^{+},\sigma_{1}^{-}\sigma_{2}^{+}] \\
&=4[\sigma_{1}^{+},\sigma_{1}^{-}]\sigma_{2}^{+}\\ &= 4\sigma_{1}^{z}\sigma_{2}^{+}~.
\end{split}
\end{equation}
Consequently
\begin{equation}
\frac{1}{4}\sigma_{1}^z[\sigma_{1}^{+},[H,\sigma_{1}^{z}]] = \frac{1}{4}\sigma_{1}^{z}4\sigma_{1}^{z}\sigma_{2}^{+} = \sigma_{2}^{+}~,
\end{equation}
as we wanted.

Carrying out the computation
\begin{equation}
\begin{split}
\sigma_{j-1}^{+}H\sigma_{j-1}^{+} &= \sigma_{j-1}^{+}(\sum_{k=1}^{N-1}2\sigma_{k}^{+}\sigma_{k+1}^{-} + 2\sigma_{k}^{-}\sigma_{k+1}^{+} + \Delta\sigma_{k}^{z}\sigma_{k+1}^{z})\sigma_{j-1}^{+}\\ &= (0 + 2\frac{I+\sigma_{j-1}^{z}}{2}\sigma_{j}^{+} + \Delta(-\sigma_{j-1}^{+})\sigma_{j}^{z})_{k=j-1}\sigma_{j-1}^{+}\\
&+\sigma_{j-1}^{+}(2\sigma_{j-2}^{+}\frac{I - \sigma_{j-1}^{z}}{2}+0+\Delta\sigma_{j-2}^{z}\sigma_{j-1}^{+})_{k+1=j-1}\\
&= (I+\sigma_{j-1}^{z})\sigma_{j}^{+}\sigma_{j-1}^{+} - \Delta\sigma_{j-1}^{+}\sigma_{j}^{z}\sigma_{j-1}^{+}\\ &+\sigma_{j-1}^{+}\sigma_{j-2}^{+}(I-\sigma_{j-1}^{z}) + \Delta\sigma_{j-1}^{+}\sigma_{j-2}^{z}\sigma_{j-1}^{+} \\ &= (I+\sigma_{j-1}^{z})\sigma_{j-1}^{+}\sigma_{j}^{+} + \sigma_{j-2}^{+}\sigma_{j-1}^{+}(I - \sigma_{j-1}^{z}) \\ &= \sigma_{j-1}^{+}\sigma_{j}^{+} + \sigma_{j-1}^{+}\sigma_{j}^{+} + \sigma_{j-2}^{+}\sigma_{j-1}^{+} - \sigma_{j-2}^{+}(-\sigma_{j-1}^{+}) \\ &= 2\sigma_{j-2}^{+}\sigma_{j-1}^{+}+ 2\sigma_{j-1}^{+}\sigma_{j}^{+}~,
\end{split}
\end{equation}
hence,
\begin{equation}
\begin{split}
[\sigma_{j-1}^{-},\sigma_{j-1}^{+}H\sigma_{j-1}^{+}] &= 2[\sigma_{j-1}^{-},\sigma_{j-2}^{+}\sigma_{j-1}^{+}]+ 2[\sigma_{j-1}^{-},\sigma_{j-1}^{+}\sigma_{j}^{+}]\\ &= 2\sigma_{j-2}^{+}(-\sigma_{j-1}^{z}) + 2(-\sigma_{j-1}^{z}\sigma_{j}^{+})\\
&= -2(\sigma_{j-2}^{+}\sigma_{j-1}^{z}+\sigma_{j-1}^{z}\sigma_{j}^{+})~,         \end{split}
\end{equation}
and so,
\begin{equation}
\begin{split}
-\frac{1}{2}\sigma_{j-1}^{z}[\sigma_{j-1}^{-},\sigma_{j-1}^{+}H\sigma_{j-1}^{+}] &= -\frac{1}{2}\sigma_{j-1}^{z}(-2)(\sigma_{j-2}^{+}\sigma_{j-1}^{z} + \sigma_{j-1}^{z}\sigma_{j}^{+})\\ &= \sigma_{j-2}^{+} + \sigma_{j}^{+}~.
\end{split}
\end{equation}

For the adjoint, we have
\begin{equation}
\begin{split}
\sigma_{2}^{-} &= (\sigma_{2}^{+})^{\dagger} = (\frac{1}{4}\sigma_{1}^{z}[\sigma_{1}^{+},[H,\sigma_{1}^{z}]])\\&= \frac{1}{4}[\sigma_{1}^{+},[H,\sigma_{1}^{z}]]^{\dagger}\sigma_{1}^{z} = \frac{1}{4}(-)[\sigma_{1}^{-},[H,\sigma_{1}^{z}]^{\dagger}]\sigma_{1}^{z} \\&= \frac{1}{4}(-)[\sigma_{1}^{-},(-)[H,\sigma_{1}^{z}]]\sigma_{1}^{z} = \frac{1}{4}[\sigma_{1}^{-},[H,\sigma_{1}^{z}]]\sigma_{1}^{z}~,
\end{split}
\end{equation}
where we used the identity
\begin{equation}
[A,B]^{\dagger} = (AB-BA)^{\dagger} = B^{\dagger}A^{\dagger} - A^{\dagger}B^{\dagger} = [B^{\dagger},A^{\dagger}] = -[A^{\dagger},B^{\dagger}]~.
\end{equation}
We also have
\begin{equation}
\begin{split}
\sigma_{j}^{-} &= (\sigma_{j}^{+})^{\dagger} = (-\sigma_{j-2}^{+}-\frac{1}{2}\sigma_{j-1}^{z}[\sigma_{j-1}^{-},\sigma_{j-1}^{+}H\sigma_{j-1}^{+}])^{\dagger}\\ &= -\sigma_{j-2}^{-} - \frac{1}{2}[\sigma_{j-1}^{-},\sigma_{j-1}^{+}H\sigma_{j-1}^{+}]^{\dagger}\sigma_{j-1}^{z} \\&-\sigma_{j-2}^{-} - \frac{1}{2}(-)[\sigma_{j-1}^{+},\sigma_{j-1}^{-}H\sigma_{j-1}^{-}]\sigma_{j-1}^{z} \\&= -\sigma_{j-2}^{-} + \frac{1}{2}[\sigma_{j-1}^{+},\sigma_{j-1}^{-}H\sigma_{j-1}^{-}]\sigma_{j-1}^{z}~,
\end{split}
\end{equation}
and with these  results we can conclude the proof.

%%%%%%%%%%%%%%%%%%%%%%%%%%%%%%%%%%%%%%%%%%%%%%%%%%%%%%%%%%%%%%%%%%%%%%%%%%%%%%%%%%%%%%%%%%%%%%%%%%%%%%%%%%%%%%%%%%%%%%%%%%%%%%%%%%%%%%%%%%%%%%%%%%%%%%%%%%

\section{Final remarks}

We believe that our results showing the general occurrence of a nontrivial
property of energy transport in quantum spin systems will enhance the interest
of researchers in quantum transport. It is worth to emphasize that the one-way
street phenomenon shown here is an effect stronger than rectification, even
a perfect rectification.

These boundary-driven quantum spin systems are the archetypal models of
nonequilibrium statistical physics, and the asymmetric versions proposed here
are not only theoretical proposals. Graded materials, for example, i.e.,
asymmetric systems with structure changing gradually in space, are abundant
in nature and can be also built. They are recurrently studied in different
areas: material science, optics, etc. An example of graded thermal diode
has been already experimentally constructed \cite{Chang}: a carbon and
boron nitride nanotube, externally coated with heavy molecules.

It is important to stress that, in particular, asymmetric versions of
$\mathit{XXZ}$ and Heisenberg chains seem to be realizable. In Refs.\cite{Endres, Barredo}, it is
shown the possibility to engineer these quantum spin Hamiltonians with
different values for the structural parameters $\alpha$ and $\Delta$.

Finally, still concerning the realizability of such systems, recent
experimental works with Rydberg atoms in optical traps \cite{Duan, Nguyen} appear associated
to Heisenberg and $\mathit{XXZ}$ models.

%%%%%%%%%%%%%%%%%%%%%%%%%%%%%%%%%%%%%%%%%%%%%%%%%%%%%%%%%%%%%%%%%%%%%%%%%%%%%%%%%%%%%%%%%%%%%%%%%%%%%%%%%%%%%%%%%%%%%%%%%%%%%%%%%%%%%%%%%%%%%%%%%%%%%%%%%%%%%%%%%%%%%%%%%%%%%%%%%%%%%%%%%%%%%%%%%%%%%%%%%%%%%%%%%%%%%%%%%%%%%%%%%%
%\newpage

\vspace*{1 cm} {\bf Acknowledgments:} This work was partially supported by CNPq (Brazil).

%%%%%%%%%%%%%%%%%%%%%%%%%%%%%%%%%%%%%%%%%%%%%%%%%%%%%%%%%%%%%%%%%%%%%%%%%%%%%%%%%%%%%%%%%%%%%%%%%%%%%%%%%%%%%%%%%%%%%%%%%%%%%%%%%%%%%%%%%%%%%%%%%%%%%%%%%%%%%%%%%%%%%%%%%%%%%%%%%%%%%%%%%%%%%%%%%%%%%%%%%%%%%%%%%%%%%%%%%%%%%%%%%%%%%%%%%%

%%%%%%%%%%%%%%%%%%%%%%%%%%%%%%%%%%%%%%%%%%%%%%%%%%%%%%%%%%%%%%%%%%%%%%%%%%%%%%%%%%%%%%%%%%%%%%%%%%%%%%%%%%%%%%%%%%%%%%%%%%%%%%%%%%%%%%%%%%%%%%%%%%%%%%%%%%

\end{document}